\documentclass[usenatbib,twocolumn,useAMS]{mn2e}
\usepackage{latexsym}
\usepackage{dcolumn}
\usepackage{bm}
\input{epsf}
\newcommand{\be}{\begin{equation}}
\newcommand{\ee}{\end{equation}}
\newcommand{\ba}{\begin{eqnarray}}
\newcommand{\ea}{\end{eqnarray}}
\newcommand{\no}{\nonumber}
\newcommand{\bi}{\begin{itemize}}
\newcommand{\ei}{\end{itemize}}
\newcommand{\bfi}{\begin{figure}
\epsfxsize=9cm
\epsffile}
\newcommand{\efi}{\end{figure}}

\newcommand{\mnras}{MNRAS}
\newcommand{\apj}{ApJ}
\newcommand{\apjl}{ApJ}

\newcommand{\aap}{AAP}
\newcommand{\prd}{PRD}

\title[The intrinsic alignment]{A proposal on the galaxy intrinsic alignment
  self-calibration in weak lensing surveys}
\author[Zhang]{Pengjie Zhang$^1$
\\$^1$Key Laboratory for Research in Galaxies and Cosmology, Shanghai
  Astronomical Observatory, Nandan Road 80, Shanghai, 200030,
  China;\\ pjzhang@shao.ac.cn}
\begin{document}
\maketitle
\begin{abstract}
The galaxy intrinsic alignment causes the galaxy ellipticity-ellipticity
power spectrum between two photometric redshifts to decrease faster with
respect to the 
redshift separation $\Delta z^P$, for fixed mean redshift. This offers a
valuable diagnosis on the intrinsic alignment. We show that the distinctive
dependences of the GG, II and GI correlations on $\Delta
z^P$ over the range $|\Delta z^P|\la 0.2$ can be understood robustly without
strong assumptions on the intrinsic alignment.   This allows us to measure the
intrinsic alignment within each conventional photo-z bin of typical size $\ga
0.2$, through lensing  tomography of photo-z bin size  $\sim 0.01$. Both the
statistical and systematical errors in the  lensing cosmology can be reduced
by this self-calibration technique. 
\end{abstract} 
\section{Introduction}
The galaxy intrinsic alignment (IA) is one of the major systematical errors
of  cosmic shear measurement. The intrinsic alignment of
physically close
galaxy pairs is correlated due to the tidal force arising from the correlated
large scale structure and thus induces the
intrinsic ellipticity-intrinsic ellipticity correlation  (the II correlation).
Due to the same reason, it is also correlated with the ambient matter
distribution, which  lenses 
background galaxies within sufficiently small angular separation and hence
couples the shape of background galaxies with 
the shape of foreground galaxies, even if the galaxy pairs are widely
separated in redshift. This induces
the gravitational 
shear-intrinsic ellipticity  correlation (the GI correlation,
\citealt{Hirata04b}).  Various
methods have been proposed  
to correct for the II correlation \citep{King02,King03,Heymans03,Takada04,Okumura09a}
and the GI correlation
\citep{Hirata04b,Heymans06,Joachimi08,Zhang08,Joachimi09,Joachimi09b,Okumura09,Kirk10,Shi10,Joachimi10}.   

Here we point out a new possibility to self-calibrate the
intrinsic alignment.  It is known that the lensing signal GG and
the contaminations II and GI have different dependences on the pair separation
$\Delta z^P$. This has motivated the proposals to reduce the II correlation
by using only cross correlation between thick photo-z bins of size $\ga 0.2$
(e.g. \citealt{Takada04}).  

What we will show in this paper is that,  these
dependences can be understood robustly 
over the range $|\Delta z^P|\la 0.2$ without heavy IA modeling.   {\it This
  allows us to self-calibrate the intrinsic alignment with basically no
  assumptions on the intrinsic 
alignment,  through  lensing
tomography of fine photo-z bin  size $\sim 0.01$ within each conventional
photo-z bin of size $\ga 0.2$}.  With this self-calibration technique,  we no
longer need to throw away the auto-correlation measurement of each
conventional thick photo-z bins. It will bring in two-fold improvement on
cosmology. First, it directly improves the cosmological constraints by
$O(10\%)$  \citep{Takada04}, since the auto-correlation measurement can now be
safely included. The price to pay is orders of magnitude more correlation
measurements due to many more photo-z bins. Second, it
extracts  valuable information on  the intrinsic alignment within each thick
photo-z bin, which helps calibrate the intrinsic alignment in the cross
correlation measurement.  This new proposal,
along with existing ones 
(e.g. \citealt{King02,Bridle07,Joachimi08,Zhang08,Bernstein09,Joachimi09,Joachimi10}),
demonstrate  
the rich   information brought in by the photo-z measurement to reduce
systematical errors and to improve cosmological constraints, which is
otherwise lost in the usual lensing tomography of thick photo-z bins.

The current proposal  is  complementary to other model independent
methods, especially  the nulling technique proposed by
\citet{Joachimi08,Joachimi09,Joachimi10}.  Our method takes advantage of the
characteristic $\Delta z^P$ dependences of the lensing signal and the intrinsic
alignment, for a fixed mean source redshift $\bar{z}^P$, to separate the
two. Over the interesting range $|\Delta z^P|\la 0.2$, since the lensing geometry
kernel is wide, the lensing signal barely changes as a function of $\Delta
z^P$ while both II and GI change significantly. On the other hand, 
the nulling technique uses the characteristic  $z^P$ dependence of the lensing
signal, arising from the lensing  geometry kernel, to suppress IA by a proper
weighting function of $z^P$.  It relies on significant variation
of the lensing kernel with respect to $z^P$ over sufficiently large redshift
range. Hence the two methods are highly complementary and we expect 
significant improvement in calibrating the intrinsic alignment by combining
the two methods.  

\section{The $\Delta z^P$ dependences}
We work on the power spectrum
$C^{\alpha\beta}(\ell,z_1^P,z_2^P)$ between a property $\alpha$ 
at photo-z $z_1^P$ and another property $\beta$ at $z_2^P$.  Here, the
superscripts (and subscripts sometime)
$\alpha,\beta=G,I,g$. ``G'' denotes the lensing convergence   $\kappa$ or the
underlying 3D matter distribution,  ``I'' the E-mode intrinsic ellipticity and
``g'', the 2D or 3D galaxy number over-density. Throughout 
this paper, we focus on the 
  $\Delta z^P\equiv z_2^P-z_1^P$ dependence with fixed multipole $\ell$ and
the mean redshift $\bar{z}^P\equiv (z_1^P+z_2^P)/2$. Thus we will use
the notation 
\be
C^{\alpha\beta}(\Delta z^P|\ell,\bar{z}^P)\equiv C^{\alpha\beta}(\ell,
z_1^P,z_2^P)
\ee
and often neglect the arguments $\ell$ and $\bar{z}^P$. 
There is an obvious symmetry $C^{\alpha\beta}(\Delta
  z^P)=C^{\beta\alpha}(-\Delta z^P)$.  It tells us $\partial
  C^{\alpha\alpha}/\partial \Delta z^P|_0=0$, a result which will become
  useful later.  Unless otherwise specified, 
  we will fix $\bar{z}^P=1.0$ and $\ell=10^3$, both are typical choices in
  weak lensing statistics.  We adopt the Limber approximation to evaluate  $C^{\alpha\beta}$ in the standard
flat
$\Lambda$CDM cosmology, 
\ba
\label{eqn:Cab}
C^{\alpha\beta}(\Delta
z^P|\ell,\bar{z}^P)&=&\frac{2\pi^2}{\ell^3}\int_0^{\infty}
\Delta^2_{\alpha\beta}\left(k=\frac{\ell}{\chi(z)},z\right)\\
&\times& W_{\alpha\beta}(z,\Delta z^P,\bar{z}^P)\chi(z)H(z)dz \ . \no
\ea
 Here, $\Delta^2_{\alpha\beta}(k,z)$ is the corresponding 3D power
spectrum variance. $\chi(z)$ and $H(z)$ are the
comoving angular diameter distance and the Hubble parameter respectively. The
weighting function
\be
W_{\alpha\beta}(z,\Delta z^P,\bar{z}^P)\equiv
W_{\alpha}(z,z_1^P)W_{\beta}(z,z_2^P)\ ,
\ee
\be
W_G(z,z^P)\equiv H^{-1}(z)\int_0^{\infty} W_L(z,z_s)p(z_s|z^P)dz_s\ ,
\ee
\be
W_I(z,z^P)=W_g(z,z^P)=p(z|z^P)\ .
\ee
Here, $W_L(z,z_s)$ is the lensing kernel for a source at $z_s$ and a lens at
$z$. $p(z|z^P)$ is the photo-z PDF, modeled as  the sum of two Gaussians
(e..g. \citealt{Ma08}),  
\ba
p(z|z^P)&=&\frac{1-p_{\rm cat}}{2\pi
  \sigma_1(z^P)}\exp\left[-\frac{(z-z^P))^2}{2\sigma^2_1(z^P)}\right]  \\
&+&\frac{p_{\rm cat}}{2\pi
  \sigma_2(z^P)}\exp\left[-\frac{(z-f_{\rm
      bias}z^P))^2}{2\sigma^2_2(z^P)}\right] \ . \no
\ea
 $p_{\rm cat}$ is the fraction of outlier galaxies, whose true redshift is
biased by a factor $f_{\rm bias}$.  We
adopt  $\sigma_{1,2}(z^P)=0.05(1+z^P)$ and $f_{\rm bias}=0.5$. This toy model
roughly
represents $p(z|z^P\sim 1)$ of a stage IV lensing survey
(e.g. \citealt{Bernstein09b}).  Stage IV dark energy surveys require $p_{\rm
  cat}<0.1\%$ \citep{Hearin10}.  However, we adopt a much more conservative 
$p_{\rm cat}=2\%$.  Due to the possible photo-z scatters, especially the
catastrophic error, $C^{GI}(\ell,z^P_1,z^P_2)$ can be non-zero even if
$z^P_1<z^P_2$ and $C^{IG}(\ell,z^P_1,z^P_2)$ can be non-zero even if $z^P_1>z^P_2$.  So the GI
contamination that we refer throughout the paper is actually  the sum of the
two, $C^{GI}+C^{IG}$. 

To proceed, we adopt the IA model of  
\citet{Schneider09} (hereafter SB09), based on the halo model prescription,
as our fiducial model. SB09 provides a fitting formula which allows
us to conveniently perform the numerical calculation.  We adopt the 
fiducial parameters of the fitting formula listed in SB09 (and also
\citealt{Kirk10}) to calculate $\Delta^2_{II}$,  
$\Delta^2_{GI}$ and hence $C^{II}$ and $C^{GI}$. 

\bfi{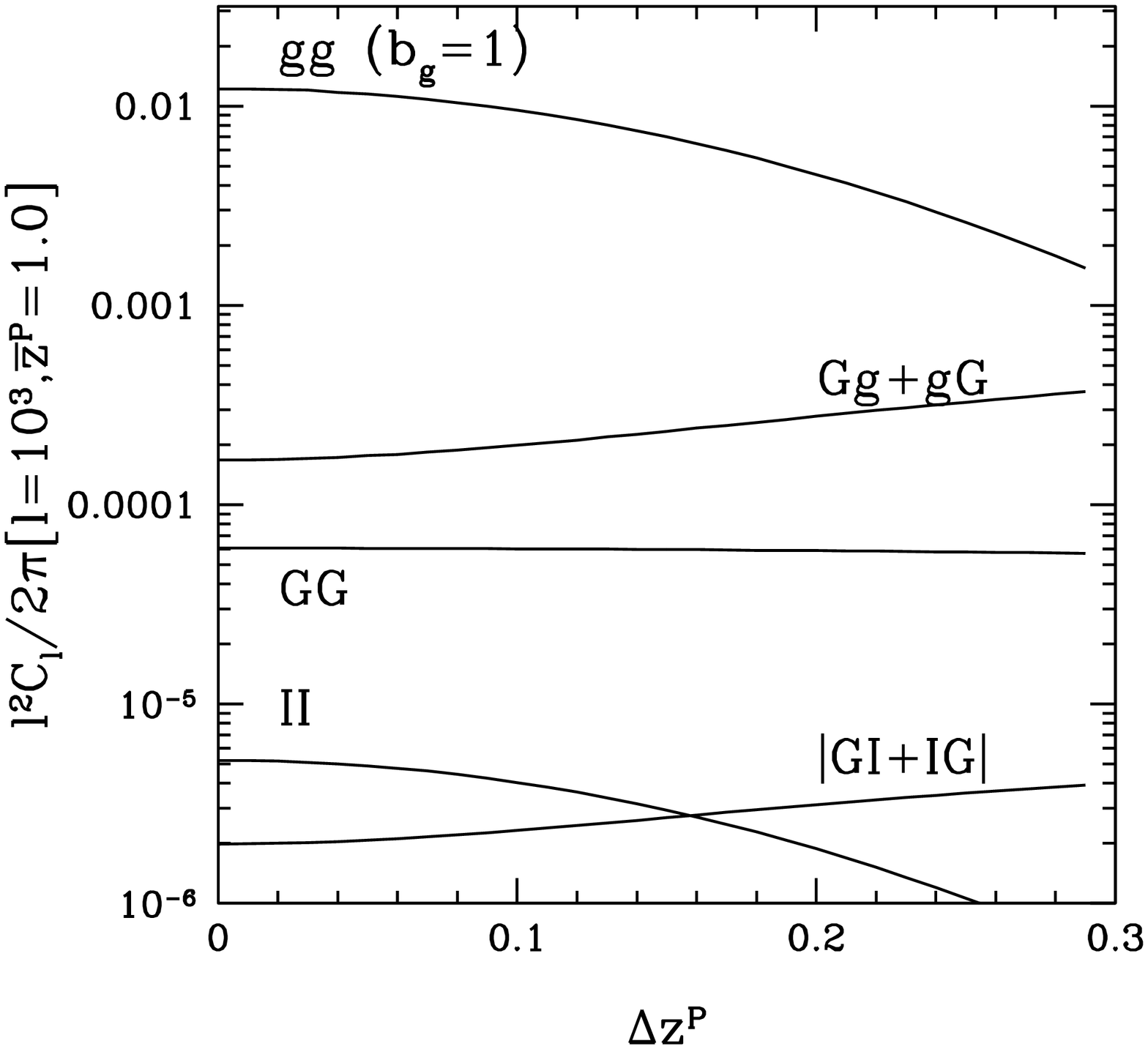}
\caption{The power spectra $C^{\alpha\beta}(\Delta
  z^P|\ell=10^3,\bar{z}^P=1)$.  We adopt the SB09 model to evaluate the II
  and GI correlations. Both II and GI  vary
  with the pair separation $\Delta z^P$ in ways significantly different from
  the lensing signal, a key for the intrinsic alignment diagnosis.  \label{fig:cl}} 
\efi

$C^{\alpha\beta}$ show distinctive dependence on
$\Delta z^P$ (Fig. \ref{fig:cl}).  
 Both II and GI  vary by $\sim 20\%$ from $\Delta z^P=0.0$ to $\Delta
z^P=0.1$ and by  $\sim 60\%$ from $\Delta z^P=0.0$ to $\Delta
z^P=0.2$. In sharp contrast,  $C^{GG}$ only decreases by $3\%$
from $\Delta z^P=0.0$ to $\Delta
z^P=0.2$.  In another word, the $\Delta z^P$ dependences of the II and GI
correlations  are more than an order of magnitude  stronger than that of the signal GG. This suggests
that, if the intrinsic alignment contaminates the galaxy
ellipticity-ellipticity power spectrum  $C^{(1)}$ by a few percent, it would
cause significantly different 
$\Delta z^P$ dependence, comparing to the case where the intrinsic alignment
is ignored, in
\ba
\label{eqn:(1)}
C^{(1)}(\Delta z^P)&=&C^{GG}(\Delta z^P)\\
&+&C^{II}(\Delta z^P)+C^{GI}(\Delta z^P)+C^{IG}(\Delta z^P)\ . \no
\ea
The II term obviously causes $C^{(1)}$ to decease faster than the case without
the II term. Since the GI term ($=C^{GI}+C^{IG}$) has a negative sign, the increment of its amplitude with
$\Delta z^P$ means that it also causes $C^{(1)}$ to decrease
faster. Thus the intrinsic alignment  always causes
$C^{(1)}$ to decrease faster with respect to $\Delta z^P$ (Fig. \ref{fig:sc}).
For example, for  the SB09 model, $C^{(1)}$ decreases by $10\%$ to $\Delta
z^P=0.2$, comparing to the $3\%$ decrease in $C^{GG}$.  

To assess the generality of the above behavior, we also investigate a
toy model. In this toy model, the intrinsic alignment has a bias (with   
respect to  matter distribution)  $b_{Im}(k,z)\propto
\left[1+\Delta^2_m(k,z)\right]^{\gamma}$ ( $\gamma\in [0,1/2]$) and the
cross correlation coefficient  $r=-1$.  Here, $\Delta^2_m(k.z)\equiv
\Delta^2_{GG}(k,z)$ is the 3D matter power spectrum.  For this set up,
$\Delta^2_{II}=b^2_{Im}\Delta^2_m$ and
$\Delta^2_{IG}=\Delta^2_{GI}=b_{Im}r\Delta^2_m=-b_{Im}\Delta^2_m$.  There is
no solid physics behind this toy model. But roughly speaking, $\gamma=0$ mimics a
class of IA models with linear dependence on the matter overdensity and
$\gamma=1/2$ with quadratic dependence.  By choosing different $\gamma$ we
can cover a wide range of IA scale and redshift dependence. Over 
$\gamma\in[0,1/2]$, we confirm the behavior that the intrinsic alignment
causes $C^{(1)}$ to decrease faster with respect to $\Delta z^P$
(Fig. \ref{fig:sc}). 

\bfi{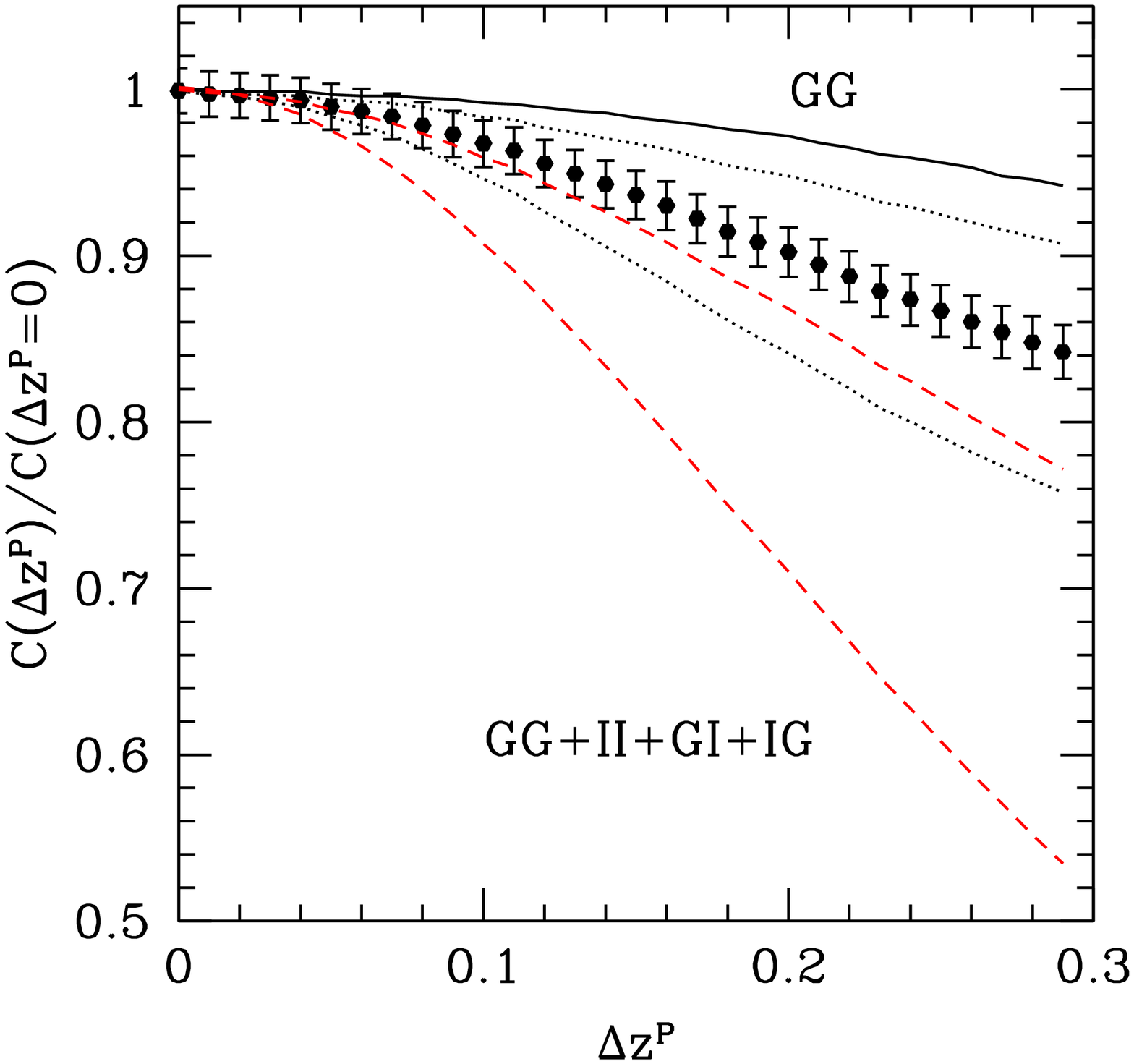}
\caption{Diagnose the intrinsic alignment through the $\Delta z^P$
  dependence of the galaxy ellipticity-ellipticity power spectrum. The data
  points  result from the fiducial 
  SB09 model and the error estimation is for LSST.  The two dot lines have
  $50\%$  weaker or stronger intrinsic alignment respectively. The
  solid line is the ideal case of no intrinsic alignment.  The two 
  dash lines are the results of an  intrinsic alignment toy model with
  $\gamma=1/2$. The intrinsic alignment of the lower one is a factor of 2
  stronger than the upper one. Results of other values of $\gamma$ have
  similar behavior. 
\label{fig:sc}} 
\efi
This faster than usual decrease is a smoking gun of the intrinsic
alignment. However, this smoking gun survives only when the photo-z
performance is reasonably good. For example, if the photo-z measurement is
completely wrong, with no correlation  with the true redshift at all, the $\Delta z^P$ dependence would
vanish. Fortunately, Fig. \ref{fig:sc}  shows  that, for typical  photo-z 
performance accessible to stage IV projects, the $\Delta z^P$ dependence is
largely preserved, even at $\Delta z^P\la 0.1$.  
 The usual lensing  tomography with coarse redshift bins
of  size $\ga 0.2$ thus misses the valuable information of the $\Delta z^P$
dependence at $\Delta z^P\la 0.2$. Such information can be
recovered by lensing tomography with fine bin size $\sim 0.01$ within each
coarse bin,  albeit requiring two orders of magnitude more computations.  

\bfi{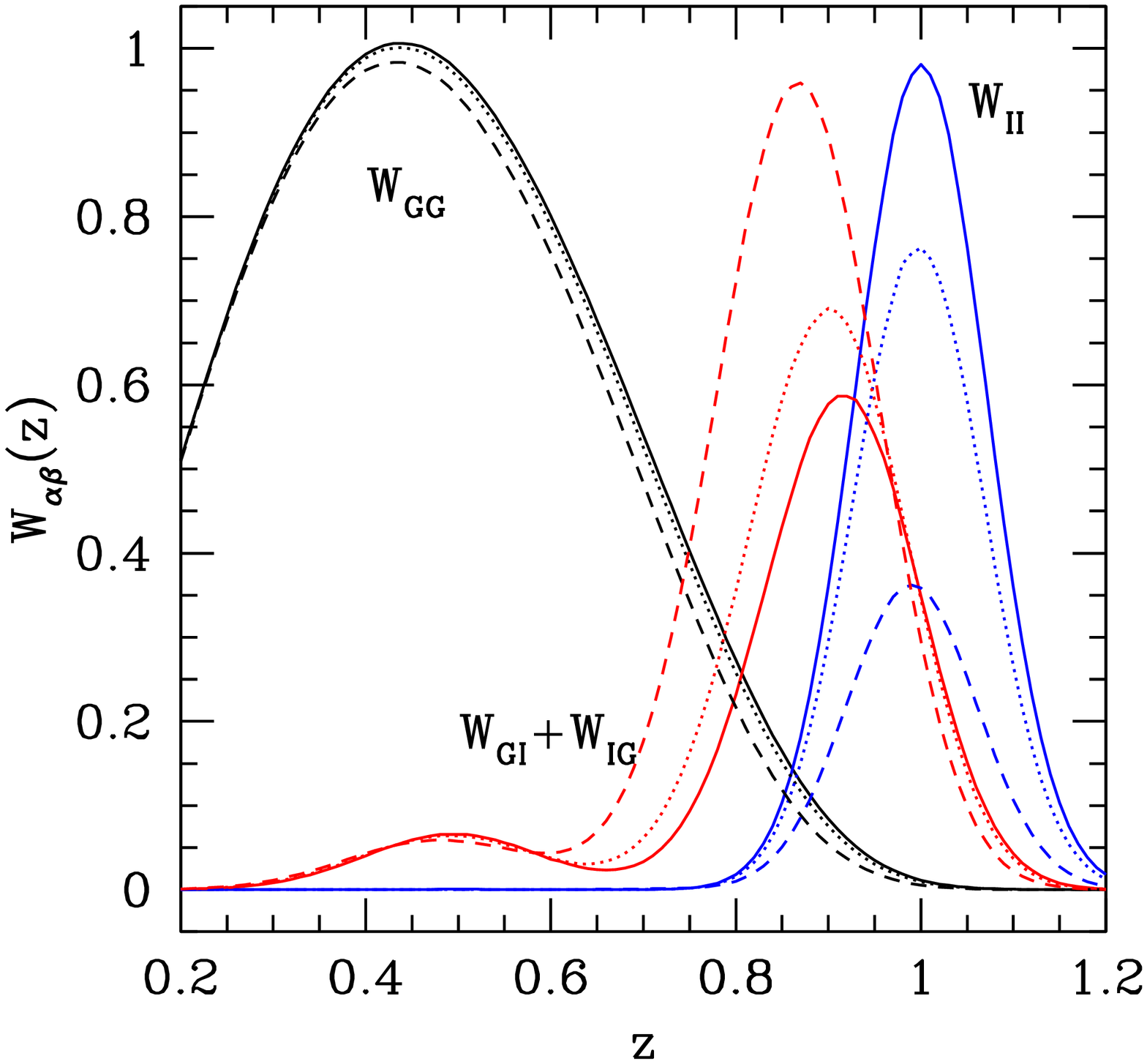}
\caption{The weighting function $W_{\alpha\beta}(z,\Delta
  z^P,\bar{z}^P=1)$. The normalizations are fixed only within each
  $\alpha\beta$.  The lines peak at left (black lines), middle (red lines)
  and right (blue) are $W_{GG}$, $W_{GI}+W_{IG}$ and $W_{II}$,
  respectively. Solid, dot and dash lines  
  have $\Delta  z^P=0.0,0.1,0.2$ respectively. The small bumps at $z\simeq
  0.5$ of $W_{GI}+W_{IG}$ are caused by the combined effect of the photo-z
  outlier at $z\sim 0.5$ and the lensing peak at $z\sim 0.5$. They have only
  weak dependence on $\Delta z^P$.\label{fig:W}} 
\efi
LSST and other surveys of comparable capability are able to measure the $\Delta
z^P$ dependence over $|\Delta z^P|\la 0.2$ to desired accuracy. To a good approximation,
$C^{\alpha\beta}$ at
different $\Delta z^P$ samples the same cosmic volume (Fig. \ref{fig:W}) and 
thus shares the same cosmic variances. The relative
differences are then  $\ll (\ell\Delta \ell f_{\rm sky})^{-1/2}=0.3\%
(10^3/\ell)^{1/2}(100/\Delta  \ell)^{1/2}f_{\rm sky}^{1/2}$, not a limiting
factor to measure the $\Delta z^P$ dependence.

The {\it random} galaxy shape shot noise dominates over the cosmic variance at
$\ell\ga 10^2$,  for a photo-z bin of  size $0.01$  at  
$\bar{z}^P=1$  with $2.5\times 10^7$ galaxies, a typical number for
LSST \citep{Zhan09}. Furthermore,  since shot noises are uncorrelated at 
different $\Delta z^P$, it becomes more difficult to measure the $\Delta z^P$
dependence. However, since the $\Delta z^P$  dependence has only weak
dependences on $\ell$ and $\bar{z}^P$, we can adopt a wide bin size $\Delta
\ell=500$ and 
average over  $10$ of these measurements of the same  $\Delta z^P$ but
different $\bar{z}^P\in [0.95,1.05]$ to beat down the shot noise to percent
level, sufficient to diagnose IA and distinguish between
some IA models of interest (Fig. \ref{fig:sc}). 

\section{Understanding the $\Delta z^P$ dependences}
Eq. \ref{eqn:Cab} shows that,  $C^{\alpha\beta}(\ell,\Delta z^P,\bar{z}^P)$ of
the same $\ell$ and $\bar{z}^P$ samples the same 3D clustering
$\Delta^2_{\alpha\beta}$, but with different weighting function
$W_{\alpha\beta}$, which is the only function in the integrand depending on
$\Delta z^P$. Hence the $\Delta z^P$ dependence in $C^{\alpha\beta}$,
especially the ratio $C(\Delta z^P)/C(\Delta z^P=0)$, should be
mainly determined by the $\Delta z^P$  dependence in $W_{\alpha\beta}$, but not $\Delta^2_{\alpha\beta}$. This 
simple fact turns out to be highly valuable. (1) It allows us to understand
the $\Delta z^P$ dependence in $C^{GG}$ to $0.1\%$ accuracy (\S
\ref{subsec:GG}).   (2) It allows us to understand the  $\Delta z^P$
dependences in $C^{II}$ (\S \ref{subsec:II}) and $C^{GI}$ (\S \ref{subsec:GI})
without strong assumptions on the intrinsic alignment, since
$W_{\alpha\beta}$ does not depend on the property of the intrinsic alignment.
These are keys towards a model-independent intrinsic alignment
self-calibration (\S \ref{sec:diagnosis}).

\subsection{The $\Delta z^P$ dependence in $C^{GG}$}
\label{subsec:GG}
$W_{GG}(z,\Delta z^P,\bar{z}^P)$ peaks at half the distance to the
source. Although the peak amplitude is sensitive to $\bar{z}^P$, it only weakly
depends on $\Delta z^P$ and only decreases by a few percent from $\Delta z^P=0$
to $\Delta z^P=0.2$. This explains the weak dependence of $C^{GG}$ on $\Delta
z^P$, which can be  well described by the Taylor
expansion around $\Delta z^P=0$ up to second order.  since $\partial
C^{GG}/\partial \Delta  z^P|_0=0$, 
\ba 
\label{eqn:AGG}
\frac{C^{GG}(\Delta z^P)}{C^{GG}(\Delta
  z^P=0)}\simeq 1-f_{GG}(\ell,\bar{z}^P)(\Delta z^P)^2\ , \\
f_{GG}(\ell,\bar{z}^P)\equiv \frac{\partial^2
C^{GG}(\Delta z^P)/\partial (\Delta z^P)^2|_0}{C^{GG}(\Delta z^P=0)}\ . \no 
\ea
We find that, $f_{GG}\in (0.5,1.0)$ for $\ell\in [20,4000]$ and $\bar{z}^P=1$.
At $\ell=10^3$, $f_{GG}=0.7$.  Eq. \ref{eqn:AGG} is accuracy to $0.1\%$ up to
$\Delta z^P=0.3$. 

\subsection{The $\Delta z^P$ dependence in the II correlation}
\label{subsec:II}
$W_{II}(z,\Delta z^P,\bar{z}^P)$  peaks sharply at a true redshift  $z_{\rm
  peak}$ (in our case, $z_{\rm peak}\simeq 1$). The peak position  is
insensitive to $\Delta z^P$, although it is
sensitive to $\bar{z}^P$, as long as the photo-z measurement is sufficiently
accurate. The 
peak amplitude decreases with  $|\Delta z^P|$ quickly,
causing the sharp decrease in $C^{II}$ at $|\Delta z^P|\gg 0.1$. This behavior
is well
known in the literature and has been applied to reduce the II correlation
by using only the cross power spectra between thick photo-z bins,
corresponding to the limit of $|\Delta z^P|\gg 0.1$.  Instead, here we explore
the II
behavior at $|\Delta z^P|\la 0.2$, namely within each thick photo-z bin of
conventional lensing tomography. 
\bfi{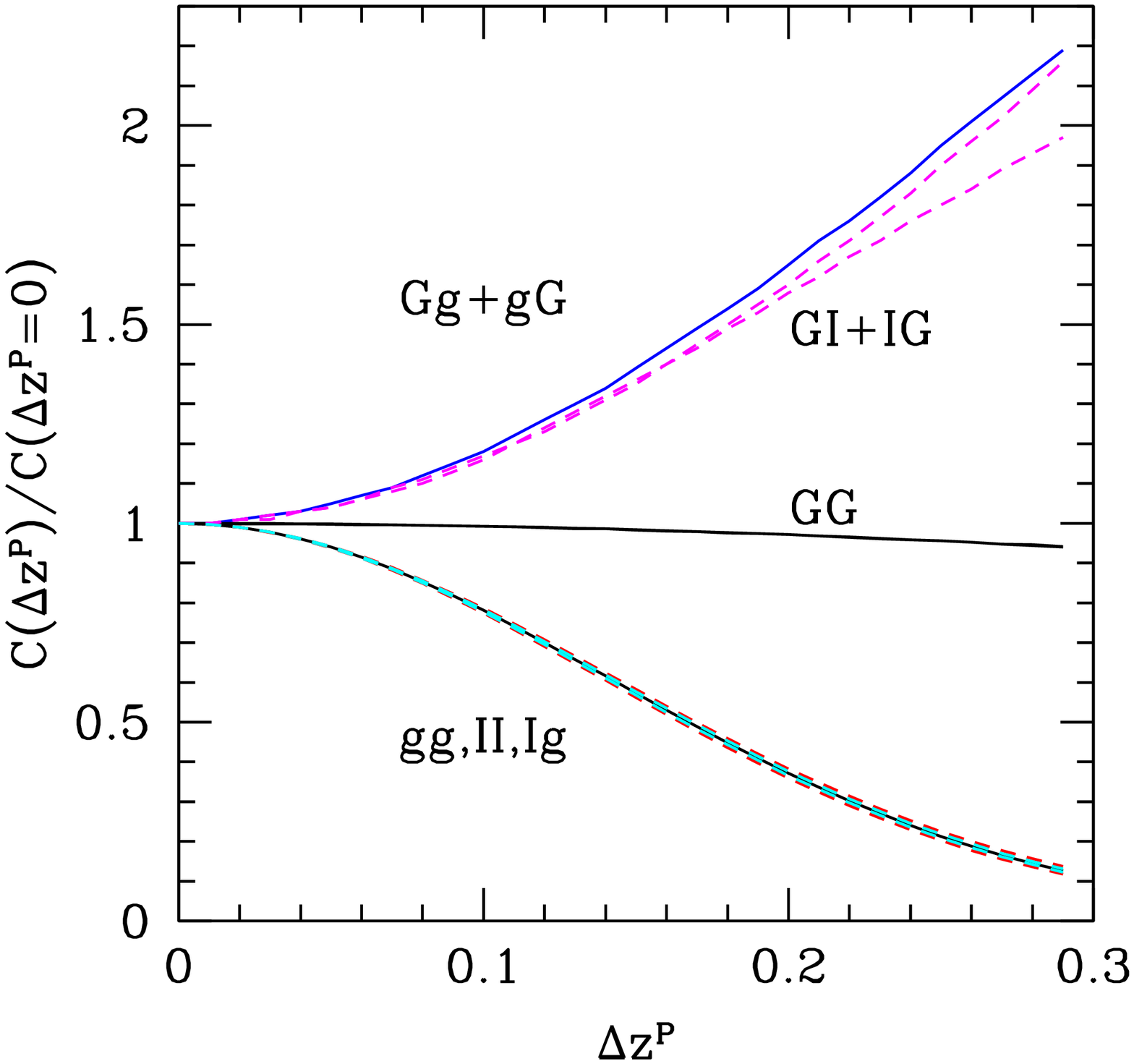}
\caption{The accuracy of the scaling relation Eq. \ref{eqn:AGG},
  \ref{eqn:AII}, \ref{eqn:AgI} \& \ref{eqn:AGI}.  The two lines labeled with
  ``GG'' are the left and right hand sides  of Eq. \ref{eqn:AGG}.  Eq. \ref{eqn:AGG}  is accurate  to   $0.1\%$ and thus the
  two lines overlap. The intrinsic alignment models demonstrated are the SB09
  model and the toy model with $\gamma=1/2$. The toy models with $\gamma<1/2$
  result in better accuracy of the scaling relation.  Eq. \ref{eqn:AII} \& \ref{eqn:AgI} 
 are  accurate to $\sim 1\%$ so the lines labeled with ``gg,II,Ig'' overlap. 
The upper solid line (blue)
  is $C^{Gg}+C^{gG}$. The two dash (magenta) lines labeled with ``GI+IG'' are the SB09
  model (lower one) and the toy model with $\gamma=1/2$ (upper one).
  Eq. \ref{eqn:AGI} is accurate   to better than  $\sim 5\%$ at $\Delta
  z^P\leq 0.2$. \label{fig:A} } 
\efi

Since $W_{II}$ has a dominant and narrow peak whose position $z_{\rm peak}$ is
insensitive to $\Delta z^P$, 
\ba
C^{II}(\Delta
z^P|\ell,\bar{z}^P)&\simeq &
\frac{2\pi^2\chi(z_{\rm peak})H(z_{\rm
peak})}{\ell^3}\Delta^2_{II}(k,z) \no \\
&\times& \int_0^{\infty} W_{II}(z,\Delta z^P,\bar{z}^P)
dz\ . 
\ea
Here, $\Delta^2_{II}(k,z)$ is evaluated at $k=\ell/\chi(z)$ and $z=z_{\rm peak}$.
The  $\Delta z^P$ dependence is completely described by the last integral. 
 In the limit that $p_{\rm cat}\rightarrow 0$ and the dependence of $\sigma_1$ 
on $z^P$ is negligible, the predicted dependence becomes exact.  Furthermore,  since
$W_{II}=W_{gg}$, we suggest the following 
relation 
\be
\label{eqn:AII}
C^{II}(\Delta z^P|\ell,\bar{z}^P)\simeq A_{II}(\ell,\bar{z}^P)C^{gg}(\Delta
z^P|\ell,\bar{z}^P)\ ,
\ee
meaning the same $\Delta z^P$ dependence in $C^{II}$ and $C^{gg}$. 
This relation  is not only  more accurate in general, but also more useful. The same
 lensing survey measures $C^{gg}(\Delta z^P)$ and thus tells the $\Delta z^P$
 dependence  in $C^{II}$, without  external knowledge on $p(z|z^P)$ nor the
 intrinsic alignment.  For basically the same reason,
 we also have  
\be
\label{eqn:AgI}
C^{Ig}(\Delta z^P|\ell,\bar{z}^P)\simeq A_{Ig}(\ell,\bar{z}^P)C^{gg}(\Delta
z^P|\ell,\bar{z}^P)\ .
\ee 
Eq. \ref{eqn:AII} \& \ref{eqn:AgI} are accurate to $1\%$, for the SB09
model and the toy model (Fig. \ref{fig:A}). To avoid  modeling
uncertainty in $A_{II}$ and $A_{Ig}$ (and $A_{GI}$ in next subsection), we
treat them as free parameters in  
the proposed self-calibration.

\subsection{The $\Delta z^P$ dependence in the GI correlation}
\label{subsec:GI}
 The corresponding weighting function for the GI
correlation ($C^{GI}+C^{IG}$) is
$W_{GI}+W_{IG}$. It peaks 
at redshift lower than $\bar{z}^P$. When $\Delta z^P$  
increases,  the true redshift separation between source and lens increases and
$W_{GI}+W_{IG}$ increases. Thus the amplitude of the GI correlation
($|C^{GI}+C^{IG}|$)  increases with $\Delta z^P$.  Due to the peak feature in
$W_{GI}+W_{IG}$, we postulate
\ba
\label{eqn:AGI}
C^{GI}(\Delta z^P|\ell,\bar{z}^P)+C^{IG}(\Delta z^P|\ell,\bar{z}^P)\simeq
A_{GI}(\ell,\bar{z}^P)\no \\
\times \left[C^{Gg}(\Delta z^P|\ell,\bar{z}^P)+C^{gG}(\Delta z^P|\ell,\bar{z}^P)\right]\ .
\ea
Since the peak is not as sharp as the one in $W_{II}$ and the peak position
does move with respect to $\Delta z^P$, and since the contribution from those
outlier galaxies at $z\sim
0.5$ becomes non-negligible (Fig. \ref{fig:W}), the accuracy of Eq. \ref{eqn:AGI} is
not as good as Eq. \ref{eqn:AII}, however, it still reaches $\sim 10\%$ up
to $\Delta z^P=0.2$ (Fig. \ref{fig:A}).  We thus take a conservative
  approach by restricting the
self-calibration technique to  $\Delta z ^P\la 0.2$. For this reason, performing
fine binning of size $0.01$ within coarse bin of size $\ga 0.2$
suffices. However, if we can improve the accuracy of Eq. \ref{eqn:AGI}
beyond $\Delta z^P=0.2$, binning the whole redshift range into fine bins of
$\Delta z^P\sim 0.01$ would be beneficial.  

The above results demonstrate that the scaling relations \ref{eqn:AII}, \ref{eqn:AgI} \&
\ref{eqn:AGI}  are indeed the manifestations of their corresponding weighting
functions,  which do not rely on the IA properties. The scale and redshift
dependences of 
the IA clustering property are significantly different between the toy
model and the SB09 model, further suggesting the generality of the above
$\Delta z^P$  dependences.  Improvements  over
\ref{eqn:AII}, \ref{eqn:AgI} \& \ref{eqn:AGI}, if necessary, can in principle
be achieved by the theory of Gaussian quadratures \citep{NR}.

\section{Self-calibrating the intrinsic alignment}
\label{sec:diagnosis}
The discovered scaling relations (Eq. \ref{eqn:AGG},
\ref{eqn:AII}, \ref{eqn:AgI} \& \ref{eqn:AGI}) can be conveniently plugged
into existing framework of  lensing tomography analysis to self-calibrate the
intrinsic alignment. Basically, lensing
surveys allow for the measurement of  the $\Delta z^P$ dependences in
$C^{(1)}$ (Eq. \ref{eqn:(1)}), $C^{(2)}$ (the ellipticity-galaxy  density
power spectrum) and 
$C^{(3)}$ (the galaxy density-galaxy density  correlation power spectrum)
\citep{Hu04,Bernstein09,Zhang08},\footnote{The magnification bias  can add non-negligible corrections
  proportional to $C^{Gg}$, $C^{Ig}$ \& $C^{GG}$ to Eq. \ref{eqn:(2)}.  Since none of
them is new unknown, it does not invalidate  the self-calibration, although
  measurement error in the galaxy luminosity function does bring new
  uncertainties. }  
\ba
\label{eqn:(2)}
C^{(2)}(\Delta z^P)&=&C^{Gg}(\Delta z^P)+C^{gG}(\Delta z^P)+2C^{Ig}(\Delta 
z^P)\ , \no \\  
C^{(3)}(\Delta z^P)&=&C^{gg}(\Delta z^P) \ .
\ea
One can show that, with the aid of Eq. \ref{eqn:AGG},
\ref{eqn:AII}, \ref{eqn:AgI} \& \ref{eqn:AGI},  measurements at 4 or more
$\Delta z^P$ allow for simultaneous reconstruction of GG, II, GI and free
parameter $f_{GG}$, $A_{GI}$, $A_{II}$ and $A_{Ig}$.  

The total contamination
(II+GI)  in $C^{(1)}$ (Eq. \ref{eqn:(1)}) can be measured with higher accuracy
than II or GI, since both causes $C^{(1)}$ to decrease faster and are thus partly
degenerate.    Basically,  a $3\%$ IA contamination in   $C^{(1)}$ would
double the $\Delta z^P$ dependence,  
observable  by LSST
(Fig. \ref{fig:sc}) or other surveys with comparable
capability.

Although we focus on $\ell=10^3$ and $\bar{z}^P=1$, the self-calibration
proposal  is
applicable to other  redshifts and angular scales.  For example, it may work
better at
larger scales $\ell\sim 10^2$, where shot noise is less an issue.  It may also
work with the presence of other errors, such as the PSF, which should have
different $\Delta z^P$ dependences. 

The intrinsic alignment may also 
induce a non-negligible B-mode shape distortion (e.g. \citealt{Heymans06}),
whose power spectrum  should have virtually the same $\Delta z^P$
dependence as $C^{gg}$ and differ significantly from  those of other B-mode
sources.  This  offers an independent way to diagnose IA from the B-mode shape
distortion measurement, although extra 
modeling is required to 
use this B-mode measurement to correct for the E-mode IA. 

The self-calibration proposal  rely
on no external measurement nor strong IA assumptions.   In this {\it letter} we
present a concept study on its feasibility.  Quantitative analysis on its performance, along with comprehensive investigation
on various complexities in realistic surveys, shall be carried out to
robustly evaluate this proposal.

{\it Acknowledgments}:  
The author thank Gary Bernstein, Scott Dodelson and Bhuvnesh Jain for useful
suggestions. This work is supported   by  the one-hundred talents program of
the Chinese Academy of Sciences (CAS), the national science
foundation of China (grant No. 10821302 \& 10973027), the CAS/SAFEA
International Partnership Program for  Creative Research Teams and the 973
program (grant No. 2007CB815401).  

\end{document}